\documentclass[twocolumn,aps,floatfix,superscriptaddress]{revtex4}
\usepackage{amsmath,amssymb,eucal,graphicx}

\begin{document}
\title{Slow Cooling of an Ising Ferromagnet}
\author{P. L. Krapivsky}
\affiliation{Department of Physics, Boston University, Boston, MA 02215, USA}

\begin{abstract} 
\noindent {\bf Abstract.}
  A ferromagnetic Ising chain which is endowed with a single-spin-flip 
  Glauber dynamics is investigated. For an arbitrary annealing protocol, we derive 
  an exact integral equation for the domain wall density. This integral equation
  admits an asymptotic solution in the limit of extremely slow cooling. For
  instance, we extract an asymptotic of the density of
  domain walls at the end of the cooling procedure when the temperature vanishes. 
  Slow annealing is usually studied using a Kibble-Zurek argument; 
  in our setting, this argument leads to approximate predictions which are 
  in good agreement with exact asymptotics. 
  
\end{abstract}

\maketitle 

\section{Introduction}

The quenching procedure that is used in numerous studies of coarsening is usually based on the following protocol: (i) Start at a high initial temperature $T_i>T_c$, where spins are disordered;  (ii) {\em Instantaneously} cool the system to a final temperature that is lower than the critical, $T_f<T_c$. The most popular choice of the initial temperature is $T_i=\infty$. Cooling to the critical temperature, $T_f=T_c$, is also often studied, especially for one-dimensional systems where $T_c=0$ or in quantum quenches. After such instantaneous quench, ordered regions begin to grow and it takes a long time (an infinite time in the thermodynamic limit) to reach the equilibrium. Understanding of the dynamics of phase-ordering kinetics, particularly the scaling regime that develops long after the quench, is the main goal \cite{B94}. 

Although the very fast cooling is natural in a number of settings, there are numerous applications (ranging from the formation of glasses to the expansion of the Universe) where slow cooling is appropriate. A popular gradual cooling scheme posits that the temperature vanishes linearly in time,
\begin{equation}
\label{slow-cool}
T(t) =T_0\cdot \left(1 - \frac{t}{\tau}\right)
\end{equation}
so as time varies in the interval $(0,\tau)$, the temperature falls from $T_i=T_0$ to $T_f=0$. To characterize the cooling scheme \eqref{slow-cool} or other gradual cooling procedure, it is natural to focus on the properties of the state at the end of the cooling procedure, particularly to find the difference between this final state [for the cooling scheme \eqref{slow-cool}, it corresponds to $t=\tau$]  and a ground state. If the cooling rate $1/\tau$ is large, the problem is not interesting as during time interval $(0,\tau)$ the system has hardly evolved. Interesting behaviors arise if cooling is very slow, $\tau \gg 1$, so that the system has evolved over long time interval $(0,\tau)$ and its state is close to a ground state. Then the problem is to describe the deviation from the ground state. 

The limit of slow cooling has been investigated in numerous papers. Earlier work has been mostly driven by applications to cosmology and proposals to study analogs of the cosmological phase transitions in the laboratory \cite{K,Z}. Recently, similar ideas and methods have been applied to the dynamics of slow quantum annealing, see e.g. \cite{ZDZ,P,D,CL}. A surprisingly little work has been done in the more standard context of thermal phase transitions. Among a few exceptions is Ref.~\cite{japan} that studies the Ising chain supplemented with Glauber's spin-flip dynamics \cite{glauber} and Ref.~\cite{france} which investigates high-dimensional systems where $T_c>0$, so that the critical temperature is passed during the quenching procedure. 

Overall, studies of slow quenching tend to rely on uncontrolled approximations. This  is not so surprising since little can be done analytically in the high-dimensional setting. Some of the uncontrolled approximations, e.g. the so-called Kibble-Zurek mechanism \cite{K,Z}, rely on sound physical ideas, yet we still want to better understand their possible limitations. Numerical (see e.g. \cite{LZ,YZ,HR}) and laboratory (see e.g. \cite{Bexp,Dexp,MPK}) experiments help in this regard, but large size simulations can be challenging (particularly for systems exhibiting quantum phase transitions), while real system always have features beyond those that are captured by theoretical models. 

Unfortunately exact results haven't been obtained even for the Ising-Glauber spin chain \cite{japan}. Here we show that this model is tractable for essentially arbitrary cooling procedure. More precisely, we establish an integral equation for the density of domain walls. This equation is {\em closed} and linear. Although it is generally unsolvable, in the interesting slow cooling limit it is possible to deduce analytically an exact asymptotic for the final density of domain walls. 

The following section \ref{cool} we write-down the governing equations for the two-spin correlation function, outline the problems in solving these equations,  and present our chief results. We then provide a complete derivation for a special cooling scheme (section \ref{SCS}).  In section \ref{general} we analyze an arbitrary cooling scheme and derive various asymptotic results announced in section \ref{cool}. Section \ref{AA} is devoted to the comparison of  exact results with approximate predictions following from the Kibble-Zurek argument. Finally in section \ref{disc} we give a summary and  briefly discuss how the cooling protocol can affect the probability for high-dimensional Ising ferromagnets to (eventually) fall into a ground state.

\section{Cooling of a Ferromagnetic Ising Chain: Main Results}
\label{cool}

In one dimension, $T_c=0$ and hence to assure coarsening we must cool the Ising chain to zero temperature: $T_f=0$. The zero-temperature evolution will continue as long as there are domain walls, so if we let the Ising chain to evolve during the time interval $\tau<t<\infty$, it will eventually reach a ground state. Therefore the density of domain walls at the end of the cooling procedure, $t=\tau$, gives the quantitative measure of the distance between the state at the end of cooling and the ground state. The deriving of this final density is our major goal. 

For concreteness, we always assume that the initial condition is spatially homogeneous. Then the chain will remain (on average) spatially homogeneous. Therefore the two-spin correlation function depends only on the separation between the spins, $G_k(t)=\langle \sigma_i(t) \sigma_{i+k}(t)\rangle$. The two-spin correlation function satisfies \cite{glauber,KRB}
\begin{equation}
\label{gk}
\frac{dG_k}{dt}=-2G_k+\gamma\left(G_{k-1}+G_{k+1}\right)
\end{equation}
for $k\geq 1$. Relation
\begin{equation}
\label{G0}
G_0(t)=\langle \sigma_i^2\rangle\equiv 1
\end{equation}
plays the role of a boundary condition. Note also that the density $\rho(t)$ of domain walls can be expressed via $G_1(t)$. Indeed,  the bond $(i,i+1)$ hosts a domain wall if $\frac{1}{2}(1-s_is_{i+1})=1$ and therefore 
\begin{equation}
\label{G1}
\rho = \left\langle \frac{1}{2}(1-\sigma_i \sigma_{i+1})\right\rangle = \frac{1-G_1}{2}
\end{equation}

The parameter $\gamma$ in \eqref{gk} is given by $\gamma = \tanh(2/T)$. [We set the ferromagnetic coupling strength $J$ to unity; otherwise we would have had 
$\gamma = \tanh(2J/T)$.] Even if we are mostly interested by the density of domain walls, there is no closed equation for this quantity. Mathematically, we need to solve an infinite set of coupled linear ordinary differential equations \eqref{gk}. Glauber solved Eqs.~\eqref{gk} in the situation when $\gamma$ is a constant, see \cite{glauber,KRB}. For the gradual cooling scheme, however, we must deal with the time-dependent parameter $\gamma(t)$. Fortunately, a lot can be done analytically.

The cooling scheme \eqref{slow-cool} is characterized by a simple linear time evolution of temperature, yet the governing equations \eqref{gk} depend on time via parameter $\gamma$ which has a more complicated time dependence: 
\begin{equation}
\label{cool-T}
\gamma=\tanh\!\left[\frac{2}{T_0}\left(1-\frac{t}{\tau}\right)^{-1}\right]
\end{equation}
As a warm-up, in the next section we shall investigate an alternative cooling scheme with a linear behavior of the parameter $\gamma$ that appears in the governing equations: 
\begin{equation}
\label{cool-linear}
\gamma(t) = t/\tau
\end{equation}
For this cooling scheme the temperature decays from infinity (at $t=0$) to zero (at $t=\tau$) according to 
\begin{equation}
\label{cool-linear-T}
T(t) = \frac{2}{\tanh^{-1}(t/\tau)}
\end{equation}
The cooling schemes \eqref{slow-cool} and \eqref{cool-linear-T} are morally equivalent. The formulas for the latter scheme are a bit simpler and for that procedure we can start with $T_i=\infty$ when initial conditions are particularly simple. 

Thus for the cooling scheme \eqref{cool-linear-T} we must solve  
\begin{equation}
\label{Gk}
\frac{dG_k}{dt}=-2G_k+\frac{t}{\tau}\left(G_{k-1}+G_{k+1}\right), \quad k\geq 1
\end{equation}
on the time interval $0\leq t\leq \tau$. The initial conditions are
\begin{equation}
\label{Gk-in}
G_k(t=0) = \delta_{k,0}
\end{equation}
and the boundary condition is \eqref{G0}. The analysis of Eqs.~\eqref{Gk} shows that in the case of slow cooling, $\tau\gg 1$,  the final density of domain walls is 
\begin{equation}
\label{DW-final}
\rho(\tau) \simeq C\,\tau^{-1/4}\,,\quad 
C=\frac{\sqrt{\pi}}{\Gamma\left(\frac{1}{4}\right)}
\end{equation}

Remarkably, an Ising chain supplemented with essentially {\em arbitrary} cooling scheme is tractable. For concreteness, we shall assume that the temperature monotonously decreases from $T_0$ to $T(\tau)=0$; then $\gamma=\gamma(t)$ is a monotonously increasing function that varies from $\gamma_0$ to $\gamma(\tau)=1$.  Actually we do not need to postulate that $\gamma(t)$ is monotonous. We consider the situation when $\gamma(t)$ varies from $\gamma_0=\tanh(2/T_0)$ to $\gamma(\tau)=1$; since $\gamma = \tanh(2/T)$, it must obey $0\leq \gamma\leq 1$, no other assumptions are required. The monotonicity of $\gamma(t)$ reflects that in typical cooling schemes the temperature is monotonously decreased, and having made such an assumption we are assured that $\gamma$ lies within the bounds $\gamma_0\leq \gamma\leq 1$. 

It turns out that in the limit of slow cooling, $\tau\gg 1$, only the behavior of 
$\gamma(t)$ near the end of the cooling procedure plays a role. For instance, in the case of the algebraic behavior,
\begin{equation}
\label{Aa}
1-\gamma(t)\simeq A\left(1-\frac{t}{\tau}\right)^\alpha \quad {\rm when} 
\quad t\to \tau, 
\end{equation}
the final density of domain walls depends only on two parameters $A$ and $\alpha$, namely 
\begin{equation}
\label{DW-gen}
\rho(\tau) \simeq C\,(2\tau)^{-\alpha/2(1+\alpha)}
\end{equation}
with amplitude
\begin{equation}
\label{amplitude}
C=\sqrt{\frac{\pi}{8}}\,\left[\Gamma\!\left(\frac{3+2\alpha}{2+2\alpha}\right)\right]^{-1}
\left[\frac{A}{1+\alpha}\right]^{\frac{1}{2(1+\alpha)}}
\end{equation}

An algebraic approach \eqref{Aa} of $\gamma(t)$ to unity is quite natural. However, for the cooling scheme \eqref{slow-cool} the behavior of $\gamma(t)$ near the end of the cooling procedure is exponential rather than algebraic
\begin{equation*}
1-\gamma(t)\simeq 2\exp\!\left\{-\frac{4/T_0}{1-t/\tau}\right\} \quad {\rm when} 
\quad t\to \tau
\end{equation*}
This suggests to analyze a family of cooling laws with an asymptotic behavior of $\gamma(t)$ of the form
\begin{equation}
\label{BBB}
1-\gamma(t)\simeq B\exp\!\left\{-\frac{b}{(1-t/\tau)^\beta}\right\} \quad {\rm when} 
\quad t\to \tau
\end{equation}
with arbitrary positive parameters $B,b,\beta$. In this situation, the asymptotic behavior of the final density of domain walls is
\begin{equation}
\label{DW-BB}
\rho(\tau) \simeq \frac{1}{4}\,\sqrt{\frac{\pi}{\tau}}
\left(\frac{\ln \tau}{b}\right)^{\frac{1}{2\beta}}
\end{equation}
This is a remarkably universal asymptotic:  The leading $\tau^{-1/2}$ algebraic factor does not depend on the parameters $B,b,\beta$ and the logarithmic prefactor depends only on one parameter $\beta$. 

We now turn to derivations of the announced results.

\section{Special Cooling Scheme}
\label{SCS}

In this section we consider the special cooling procedure \eqref{cool-linear}. First, we employ an exact analysis and then turn to the asymptotic behavior of the final density of domain walls.

\subsection{Exact Analysis}
\label{EA}

To handle an infinite set of ordinary differential equations \eqref{Gk} we recast it into a single differential equation using a generating function technique. Multiplying \eqref{Gk} by $z^k$ and summing over all $k\geq 1$ we find that the generating function 
\begin{equation}
G(z,t) = \sum_{k\geq 1} G_k(t)\,z^k
\end{equation}
satisfies
\begin{equation}
\label{G-eq}
\frac{\partial G}{\partial t} = 
\left[-2+\frac{t}{\tau}\left(z+z^{-1}\right)\right]G + 
\frac{t}{\tau}\left[z-G_1(t)\right]
\end{equation}
Solving \eqref{G-eq} subject to $G(z,t=0) = 0$, we get
\begin{equation*}
G = \int_0^t dt'\,\frac{t'}{\tau}\left[z-G_1(t')\right]\exp\!\left[2(t'-t)+
\eta\,\frac{z+z^{-1}}{2}\right]
\end{equation*}
where $\eta=(t^2-t'^2)/\tau$. Recalling that the exponential factor $\exp\!\left[\eta\,\frac{z+z^{-1}}{2}\right]$ is the generating function for the modified Bessel functions 
\begin{equation*}
\exp\!\left[\eta\,\frac{z+z^{-1}}{2}\right] = \sum_{n=-\infty}^\infty z^n\,I_n(\eta)
\end{equation*}
we conclude that
\begin{equation}
\label{Gk-sol}
G_k(t) = \int_0^t dt'\,\frac{t'}{\tau}\,e^{2(t'-t)}\left[I_{k-1}(\eta)-G_1(t')I_k(\eta)\right]
\end{equation}

Thus to determine $G_1$ we need to solve an integral equation 
\begin{equation}
\label{G1-eq}
G_1(t) = \int_0^t dt'\,\frac{t'}{\tau}\,e^{2(t'-t)}\left[I_0(\eta)-G_1(t')I_1(\eta)\right]
\end{equation}
Higher correlation functions $G_k$ with $k\geq 2$ are then expressed via $G_1$ and \eqref{Gk-sol}. 

The above results are exact. Now we turn to the most interesting case of slow cooling, $\tau\gg 1$, and compute the final density of domain walls.

\subsection{Final Density of Domain Walls}
\label{FDDW}

Using \eqref{G1}, we re-write the integral equation  \eqref{G1-eq} in terms of the density of domain walls:
\begin{eqnarray}
\label{DW-eq}
1&-& \int_0^t dt'\,\frac{t'}{\tau}\,e^{2(t'-t)}\left[I_0(\eta)-I_1(\eta)\right]\nonumber\\
 &=&2\rho(t)+2\int_0^t dt'\,\frac{t'}{\tau}\,e^{2(t'-t)}\,I_1(\eta)\,\rho(t')
\end{eqnarray}
We are mostly interested in the final density $\rho(\tau)$. Hence we specialize \eqref{DW-eq} to $t=\tau$. It is also convenient to change the integration variable, $t'\to \eta=(\tau^2-t'^2)/\tau$. We arrive at the integral equation
\begin{eqnarray}
\label{DW-final-eq}
1&-&\frac{1}{2} \int_0^\tau d\eta\,\mathcal{E}(\eta,\tau)\left[I_0(\eta)-I_1(\eta)\right]\nonumber\\
 &=&2\rho(\tau)+\int_0^\tau d\eta\,\mathcal{E}(\eta,\tau)\,I_1(\eta)\,\rho\left(\tau\sqrt{1-\eta/\tau}\right)
\end{eqnarray}
where we used the shorthand notation
\begin{equation}
\label{E-def}
\mathcal{E}(\eta,\tau) \equiv \exp\!\left[2\tau\left(\sqrt{1-\eta/\tau} -1\right)\right]
\end{equation}

The chief contribution to the integral on the right-hand side of  \eqref{DW-final-eq} is gathered when $\eta\sim \sqrt{\tau}$ [see Eq.~\eqref{RHS} below].  Therefore we can replace $\rho\left(\tau\sqrt{1-\eta/\tau}\right)$ by $\rho(\tau)$ and the exponential factor $\mathcal{E}(\eta,\tau)$ by 
\begin{equation}
\mathcal{E}(\eta,\tau) = e^{-\eta-\eta^2/4\tau}
\end{equation}
as it follows by expanding the term inside the square brackets in Eq.~\eqref{E-def}. Thus the integral on the right-hand side of \eqref{DW-final-eq} becomes
\begin{equation}
\label{int-RHS}
\rho(\tau)\int_0^\tau d\eta\,e^{-\eta-\eta^2/4\tau}\,I_1(\eta)
\end{equation}
We can further use the asymptotic formula for the Bessel function,
\begin{equation}
\label{Bess-ass}
I_1(\eta)\simeq \frac{e^\eta}{\sqrt{2\pi\eta}}\quad {\rm when}\quad \eta\gg 1,
\end{equation}
and replace the upper limit in \eqref{int-RHS} by infinity. Thus we obtain
\begin{equation}
\label{RHS}
\rho(\tau)\int_0^\infty d\eta\,e^{-\eta^2/4\tau}\,(2\pi\eta)^{-1/2}
= \rho(\tau)\,\frac{\Gamma\left(\frac{1}{4}\right)\,\tau^{1/4}}{\sqrt{4\pi}}
\end{equation}
This obviously dominates the second term $2\rho(\tau)$ on the right-hand side of  \eqref{DW-final}. 

On the left-hand side of  \eqref{DW-final} both terms are comparable. The integral 
\begin{equation*}
\int_0^\tau d\eta\,\mathcal{E}(\eta,\tau)\left[I_0(\eta)-I_1(\eta)\right]
\end{equation*}
is gathered in the region $\eta=\mathcal{O}(1)$. Thus we can replace $\mathcal{E}(\eta,\tau)$ by $e^{-\eta}$ and integrate up to infinity. Using 
\begin{equation}
\label{Bess-int}
\int_0^\infty d\eta\,e^{-\eta}\left[I_0(\eta)-I_1(\eta)\right] = 1
\end{equation}
we conclude that the left-hand side of  \eqref{DW-final-eq} is asymptotically equal to $1-1/2=1/2$. Equating this to the asymptotic expression \eqref{RHS} for the right-hand side we arrive at the announced results \eqref{DW-final} for the final density of domain walls.

\section{General Cooling Procedure}
\label{general}

For a cooling scheme characterized by an arbitrary function $\gamma(t)$, the solution is found using the same procedure as in Sect.~\ref{EA}. Instead of the integral equation \eqref{G1-eq} we get
\begin{equation}
\label{G1-eq-gen}
G_1(t) = \int_0^t dt'\,\gamma(t')\,e^{2(t'-t)}\left[I_0(\eta)-G_1(t')I_1(\eta)\right]
\end{equation}
with $\eta = 2\int_{t'}^t dt''\,\gamma(t'')$. The final density of domain walls satisfies
\begin{eqnarray}
\label{DW-final-gen}
1&-&\frac{1}{2} \int_0^\zeta d\eta\,e^{2(t'-\tau)}\left[I_0(\eta)-I_1(\eta)\right]\nonumber\\
 &=&2\rho(\tau)+\int_0^\zeta d\eta\,e^{2(t'-\tau)}\,I_1(\eta)\,\rho(t')
\end{eqnarray}
Note that $t'$ that appears in $\rho(t')$ and $e^{2(t'-\tau)}$ should be expressed via 
$\eta$ by inverting 
\begin{equation}
\eta(t') = 2\int_{t'}^\tau dt''\,\gamma(t'')
\end{equation}
Further, the upper limit in integrals in Eq.~\eqref{DW-final-gen} is given by $\zeta=\eta(t'=0) = 2\int_0^\tau dt''\,\gamma(t'')$. 

The dominant contributions to both integrals in \eqref{DW-final-gen} are gathered in the region where $1-t'/\tau\ll 1$. In this region the relation between $t'$ and $\eta$ simplifies. We now perform the calculations for two families of cooling procedures introduced in Sect.~\ref{cool}. 

\subsection{Algebraic Cooling}
\label{AC}

For the family of cooling procedures \eqref{Aa} we have
\begin{eqnarray*}
\eta &=& 2\int_{t'}^\tau dt''\,\gamma(t'')\\
&=& 2(\tau-t') - \frac{A}{1+\alpha}\,\tau\left(1-\frac{t'}{\tau}\right)^{1+\alpha}+\ldots
\end{eqnarray*}
or equivalently
\begin{equation*}
1-\frac{t'}{\tau} = \frac{\eta}{2\tau}
+\frac{A}{1+\alpha}\,\left(\frac{\eta}{2\tau}\right)^{1+\alpha}+\ldots
\end{equation*}
which leads to 
\begin{equation}
\label{Ett}
e^{2(t'-\tau)} = \exp\!\left[-\eta 
- \frac{A}{1+\alpha}\,\frac{\eta^{1+\alpha}}{(2\tau)^\alpha}\right]
\end{equation}
Using \eqref{Ett} and \eqref{Bess-ass} we find that the integral on the right-hand side of Eq.~\eqref{DW-final-gen} is asymptotically 
\begin{equation}
\label{int-Gamma}
\rho(\tau)\int_0^\infty \frac{d\eta}{\sqrt{2\pi\eta}}\,
 \exp\!\left[-\frac{A}{1+\alpha}\,\frac{\eta^{1+\alpha}}{(2\tau)^\alpha}\right]
\end{equation}
Computing the integral in \eqref{int-Gamma} we find that the asymptotic of the right-hand side of Eq.~\eqref{DW-final-gen} is given by
\begin{equation*}
{\rm RHS} = \sqrt{\frac{2}{\pi}}\,
\Gamma\!\left(\frac{3+2\alpha}{2+2\alpha}\right)
\left(\frac{1+\alpha}{A}\right)^{\frac{1}{2(1+\alpha)}}
(2\tau)^{\frac{\alpha}{2(1+\alpha)}}\rho(\tau)
\end{equation*}
The left-hand side of Eq.~\eqref{DW-final-gen} is asymptotically
\begin{equation*}
{\rm LHS} = 1-\frac{1}{2} \int_0^\infty d\eta\,e^{-\eta}\left[I_0(\eta)-I_1(\eta)\right]
= \frac{1}{2}
\end{equation*}
where in the last step we have used \eqref{Bess-int}. Equating the RHS to the LHS we arrive at \eqref{DW-gen}--\eqref{amplitude}.

\subsection{Exponential Cooling}
\label{EC}

For the family of cooling procedures \eqref{BBB} we have
\begin{equation*}
\eta=2(\tau-t')-2B\int_{t'}^\tau dt\,\exp\!\left\{-\frac{b}{(1-t/\tau)^\beta}\right\}
\end{equation*}
Computing the integral and inverting we find (keeping two leading terms)
\begin{equation*}
1-\frac{t'}{\tau} = \frac{\eta}{2\tau}
+\frac{B}{b\beta}\,\left(\frac{\eta}{2\tau}\right)^{1+\beta}
\exp\!\left\{-b\left(\frac{\eta}{2\tau}\right)^{-\beta}\right\}
\end{equation*}
Hence the exponential factor in the integrands in \eqref{DW-final-gen} is
\begin{equation*}
e^{2(t'-\tau)} = \exp\!\left[-\eta 
- \frac{B}{b\beta}\,\left(\frac{\eta}{2\tau}\right)^{\beta}\eta\,
\exp\!\left\{-b\left(\frac{\eta}{2\tau}\right)^{-\beta}\right\}\right]
\end{equation*}
and therefore the RHS of \eqref{DW-final-gen} is asymptotically 
\begin{equation*}
\rho(\tau)\int_0^\infty \frac{d\eta}{\sqrt{2\pi\eta}}\,\exp\!\left[
- \frac{B}{b\beta}\,\left(\frac{\eta}{2\tau}\right)^{\beta}\eta\,
\exp\!\left\{-b\left(\frac{\eta}{2\tau}\right)^{-\beta}\right\}\right]
\end{equation*}
The integral has a simple asymptotic behavior. Indeed, changing $\eta\to \xi$ via
$\eta = 2\tau(\xi/\ln \tau)^{1/\beta}$ we recast the exponential factor in the integrand into 
\begin{equation}
\label{exp-factor}
\exp\!\left[-\frac{2B}{b\beta}\,\left(\frac{\xi}{\ln \tau}\right)^{1+1/\beta}
\tau^{1-b/\xi}\right]
\end{equation}
In the $\tau\to\infty$ limit, the exponential factor \eqref{exp-factor} is asymptotically 
equal to 1 when $\xi<b$ and zero when $\xi>b$. Therefore the RHS is asymptotically
\begin{equation}
\label{RHS-EC}
\rho(\tau)\int_0^{2\tau(b/\ln\tau)^{1/\beta}} \frac{d\eta}{\sqrt{2\pi\eta}}
= 2\rho(\tau) \sqrt{\frac{\tau}{\pi}\left(\frac{b}{\ln \tau}\right)^{1/\beta}}
\end{equation}
The LHS of \eqref{DW-final-gen} is asymptotically 1/2 as before. Equating it to \eqref{RHS-EC} we arrive at the announced result \eqref{DW-BB}.

\section{Approximate Analysis}
\label{AA}

A Kibble-Zurek (KZ) argument has been proposed \cite{K,Z} to determine the final density of  defects (domain walls in our case). The KZ argument leads to qualitatively correct results, yet it involves an uncontrolled approximation and cannot predict more subtle quantitative features, e.g. amplitudes in the scaling laws. In most examples, it is impossible to estimate the quantitative error as we do not possess (asymptotically) exact results. The ferromagnetic Ising chain provides a rare exception and hence it allows us to probe the quality of an approximation provided by the KZ argument. 

The KZ argument suggests to estimate $\rho(\tau)$ in the following way:
\begin{enumerate}
\item During the initial regime the density of domain walls $\rho(t)$ is assumed to be given by its equilibrium value at the corresponding temperature, $\rho(t)=\rho_{\rm eq}[T(t)]$. This assumption is natural since the cooling rate is very slow.
\item During the final regime the density of domain walls $\rho(t)$ is assumed to be constant, $\rho(t)=\rho_*$ where $\rho_*=\rho_{\rm eq}(T_*)$ with $T_*=T(t_*)$. This freezing also seems plausible as the density $\rho_*$ is very small and during the final regime the density should not decrease much.
\item The Ising model endowed with a Glauber dynamics is characterized by the typical time $r_{\rm eq}(T)$ to relax to equilibrium. We now postulate that the separation time is determined from the criterion $r_{\rm eq}(T_*)=\tau-t_*$. 
\end{enumerate}

First, we need to find the equilibrium solution. In other words, we must solve the stationary version of Eqs.~\eqref{gk}. These equations admit an exponential equilibrium solution: By inserting $G_k=\eta^k$ into \eqref{gk} we find that the right-hand side vanishes when $2=\gamma(\eta+\eta^{-1})$. {}From this relation, 
$\eta=\gamma^{-1}\left[1-\sqrt{1-\gamma^2}\right]$, and therefore
\begin{equation}
\rho_{\rm eq}=\frac{1-\eta}{2}=\frac{1}{2}\left[1-\frac{1-\sqrt{1-\gamma^2}}{\gamma}\right]
\end{equation}

The relaxation time is to a degree the matter of agreement. Equation \eqref{gk} suggests that 
\begin{equation}
\label{rel}
r_{\rm eq} = \frac{1}{2[1-\gamma]} 
\end{equation}
as a reasonable definition. Combining \eqref{rel} and the criterion $r_{\rm eq}(T_*)=\tau-t_*$ we arrive at 
\begin{equation}
\label{crit}
 \frac{1}{2[1-\gamma_*]} =  \tau-t_*\,, \quad \gamma_*=\gamma(t_*)
\end{equation}

Let us begin with the special cooling procedure \eqref{slow-cool}. Then $\gamma_*=t_*/\tau$ and therefore \eqref{crit} gives $1-\gamma_*=1/\sqrt{2\tau}$.  Hence
\begin{equation}
\label{DW-final-KZ}
\rho_* = \frac{1}{2}\left[1-\frac{1-\sqrt{1-\gamma_*^2}}{\gamma_*}\right]
\simeq 2^{-3/4}\tau^{-1/4}
\end{equation}
Comparing \eqref{DW-final} and \eqref{DW-final-KZ} we see that the exact amplitude
$C=\pi^{1/2}/\Gamma(1/4)\Doteq 0.48887$ is smaller than the prediction from the KZ argument $C_{\rm KZ}=2^{-3/4}\Doteq 0.59460$.

\subsection{Algebraic Cooling}

For the family of cooling schemes \eqref{Aa} we recover the correct scaling law \eqref{DW-gen}, but with a slightly erroneous amplitude 
\begin{equation}
\label{amplitude-KZ}
C_{\rm KZ}=2^{-1/2}\,A^{1/(2+2\alpha)}
\end{equation}
Interestingly, the KZ argument correctly predicts even the dependence on $A$: The ratio of the exact and approximate amplitudes is the function of $\alpha$ alone
\begin{equation}
\label{CC}
\frac{C}{C_{\rm KZ}}=\frac{\sqrt{\pi}}{2}\,\left[\Gamma\!\left(\frac{3+2\alpha}{2+2\alpha}\right)\right]^{-1}
\left[\frac{1}{1+\alpha}\right]^{\frac{1}{2(1+\alpha)}}
\end{equation}
This ratio approaches to 1 as $\alpha\to 0$ and to $\sqrt{\pi/4}\Doteq 0.8862269$ when $\alpha\to\infty$. The largest deviation from unity is $0.7913\ldots$, so the maximal quantitative mistake of the KZ argument is about 20\%. 

\subsection{Exponential Cooling}

For the family of cooling procedures \eqref{BBB} we find
\begin{equation}
\label{DW-BB-KZ}
\rho_* \simeq \frac{1}{\sqrt{4\tau}}
\left(\frac{\ln \tau}{b}\right)^{\frac{1}{2\beta}}
\end{equation}
Note that the ratio of the asymptotically exact density given by \eqref{DW-BB} to the prediction \eqref{DW-BB-KZ} of the KZ argument is equal to 
$\sqrt{\pi/4}$; it is a universal number as it does not depend on the parameters $B,b,\beta$ of the cooling procedures \eqref{BBB}. The same ratio characterizes algebraic cooling schemes \eqref{Aa} in the $\alpha\to\infty$ limit. This seems to be a general feature: If $\gamma(t)$ approaches to unity faster than any power law, then the KZ argument quickly gives a prediction and the exact answer is just $\sqrt{\pi/4}$ times smaller. 

\section{Discussion}
\label{disc}

This work was inspired by the desire to test the validity of the Kibble-Zurek argument for the spin chain supplemented with non-conservative (Glauber) dynamics. The conclusion is that in this setting, the KZ argument is very good, namely it gives correct exponents and even the amplitudes are approximated reasonably well. The chief outcome of this work is actually unrelated to the KZ argument or any other approximation, namely we have shown that the Ising-Glauber chain remains solvable in the general case of an arbitrary (externally imposed) temperature drive $T=T(t)$. The term `solvable' here means that for the most interesting characteristic, the density of domain walls, we have found a closed linear inhomogeneous integral equation. It does not seem feasible to express a solution in quadratures, but interesting asymptotically exact results have been extracted. We have also given expressions of other two-spin correlations functions via the density of domain walls. 

What would happen if we shall continue beyond the point $t=\tau$, that is, we shall follow the zero-temperature evolution for $t>\tau$.  In one dimension, domain walls will continue to diffuse and annihilate upon collisions; eventually the system will reach a ground state. In the case of instantaneous cooling, higher-dimensional behaviors are different:  In two dimensions, the Ising ferromagnet may not reach a ground state (it can get trapped in a metastable stripe state); in three dimensions, a ground state is never reached \cite{victor}. It seems plausible that for slow cooling the system may not get trapped, or the probability of getting trapped decreases as $\tau\to\infty$. This appears an interesting challenge and in two dimensions at least, one may even hope for theoretical progress \cite{kip}. 

\section*{Acknowledgments}

I am grateful to Leticia Cugliandolo for discussions that revitalized my interest to the cooling problem. This work has been supported by NSF grant CCF-0829541.

\end{document}